# Optical imaging of strain in two-dimensional crystals

Lukas Mennel*, Marco M. Furchi*, Stefan Wachter, Matthias Paur, Dmitry K. Polyushkin, and Thomas Mueller‡

*Vienna University of Technology, Institute of Photonics, Gußhausstraße 27-29, 1040 Vienna, Austria*

\* These authors contributed equally to this work.
‡ Corresponding author: thomas.mueller@tuwien.ac.at

**Strain engineering is widely used in material science to tune the (opto-)electronic properties of materials and enhance the performance of devices. Two-dimensional atomic crystals are a versatile playground to study the influence of strain, as they can sustain very large deformations without breaking. Various optical techniques have been employed to probe strain in two-dimensional materials, including micro-Raman and photoluminescence spectroscopy. Here we demonstrate that optical second harmonic generation constitutes an even more powerful technique, as it allows to extract the full strain tensor with a spatial resolution below the optical diffraction limit. Our method is based on the strain-induced modification of the nonlinear susceptibility tensor due to a photoelastic effect. Using a two-point bending technique, we determine the photoelastic tensor elements of molybdenum disulfide. Once identified, these parameters allow us to spatially image the two-dimensional strain field in an inhomogeneously strained sample.**

## INTRODUCTION

The properties of materials can be strongly influenced by strain. For example, local straining techniques are employed in modern silicon field-effect transistors to reduce the carrier effective mass and increase the carrier mobility [1]. In optoelectronics, strain can be used to transform indirect band gap semiconductors into direct gap materials with strongly enhanced radiative efficiencies, or to tune the emission wavelength of light emitters [2-5]. While silicon typically breaks at strain levels of ~1.5 %, two-dimensional (2D) atomic crystals [6] can withstand strain of >10 % [7, 8], making them promising candidates for stretchable and flexible electronics [9]. Their high flexibility further also allows for folding or wrapping them around (lithographically defined) nanostructures to induce spatially inhomogeneous atom displacements. This provides an opportunity to modify their electronic structure such that excitons can be funneled into a small region of a layered semiconductor [10,11] to form exciton condensates, improve solar cell efficiencies, or realize single photon emitters [12-15].

In order to better understand and improve the performance of devices, analytical tools are required that are capable of noninvasive strain imaging at the submicron scale. Traditionally, X-ray diffraction (XRD) is employed, but submicron spatial resolution can in most cases only be achieved by the use of coherent radiation, e.g. from a synchrotron [16]. High-resolution transmission electron microscopy (TEM) based techniques [17] rely on precise measurements of atom column positions, but they offer only a small field of view and are invasive, as they require thin specimens. Electron backscatter diffraction (EBSD) techniques [18] can overcome some of those limitations, but require flat surfaces and comparison of the experimental data with simulations to analyze the complex diffraction patterns. Optical techniques, such as micro-Raman [19-21] and photoluminescence (PL) [10, 22, 23] spectroscopy, offer an interesting alternative. They are noninvasive, allow for large-area imaging with submicron spatial resolution and are simple to set up.

Strain-induced second harmonic generation (SHG) has been investigated for decades [24] and has been employed to generate frequency-doubled light in centrosymmetric materials, such as silicon [25–30]. The second harmonic response in these works is typically described by a phenomenological modification of the nonlinear susceptibility of the unstrained material [31-35]. In a more thorough theoretical investigation, Lyubchanskii *et al.* demonstrated that strain and nonlinear susceptibility are connected via a photoelastic tensor [36, 37].

Here, we adapt this theory and determine, to our knowledge for the first time, all photoelastic tensor elements of a material from SHG. Once identified, these parameters allow us to spatially map the full strain tensor in a mechanically deformed 2D material with a spatial resolution below the diffraction limit of the excitation light. As a representative example, we present results obtained from $MoS_2$ – a layered transition metal dichalcogenide (TMD) semiconductor [38, 39]. Our technique, however, is not only limited to 2D materials, but is applicable to any thin crystalline film. It establishes a novel optical strain probing technique that provides an unprecedented depth of information.

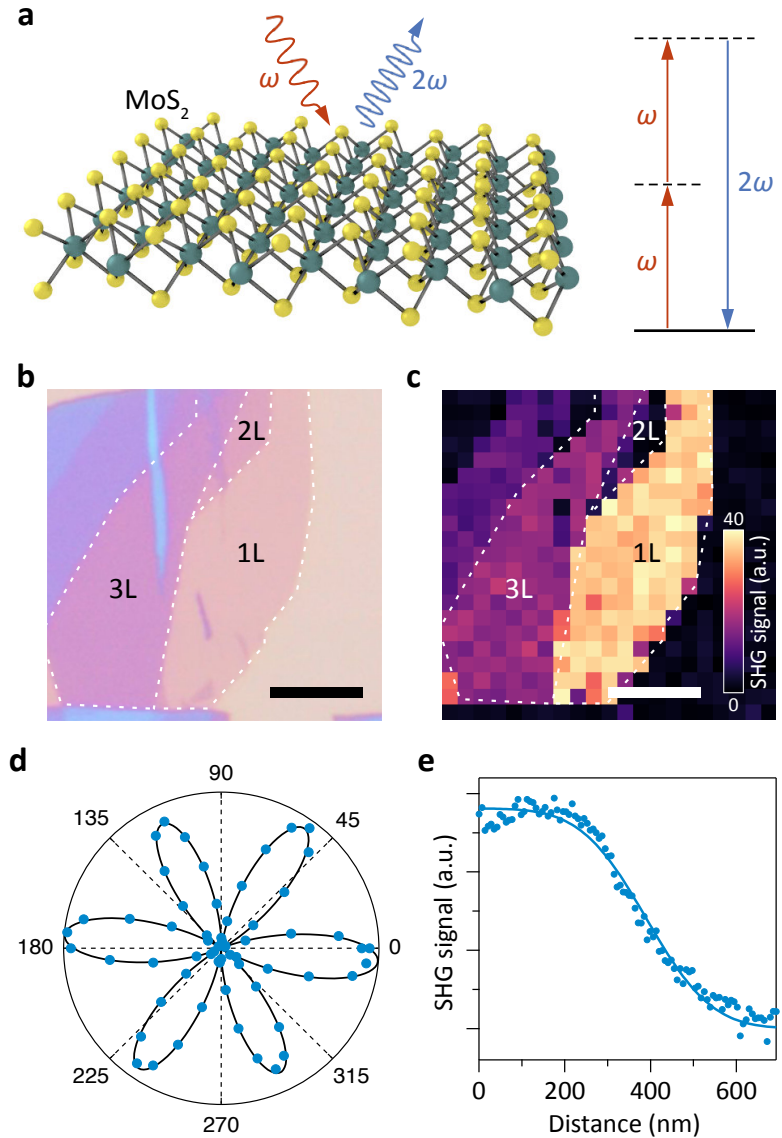

**Figure 1 | Second harmonic generation in MoS₂. a,** Schematic illustration of the SHG process. **b,** Micrograph of a mechanically exfoliated MoS₂ flake on a Si/SiO₂ wafer, and **c,** corresponding SHG amplitude. Note that the SHG signal is absent in the bilayer region. Scale bars, 5 µm. **d,** Linear polarization dependence of the SHG intensity from an MoS₂ monolayer. **e,** SHG line scan across an MoS₂ edge. Symbols: experimental data. Line: fit of a function of the form $1 - \mathrm{erf}(\sqrt{2}x/w)$, where erf() is the error function and $w$ is the waist, from which a spatial resolution of 280 nm (FWHM) is determined.

## RESULTS

**Theoretical description.** SHG is a nonlinear optical process in which two photons with the same frequency $\omega$ combine into a single photon with double frequency, as schematically depicted in Figure 1a. As only non-centrosymmetric crystals possess a second-order nonlinear susceptibility, SHG in 2H-stacked TMDs requires odd layer thickness [31-35].

Figure 1b shows a microscope image of a mechanically exfoliated MoS$_2$ flake on a Si/SiO$_2$ (280 nm) wafer. The corresponding SHG signal amplitude in Figure 1c shows the expected layer dependence and quadratic scaling behavior with excitation power (see Supplementary Figure 2). The polarization dependence of the SHG intensity, plotted in Figure 1d, reflects the underlying D$_{3h}$ symmetry of the TMD crystal [31-35]. The spatial resolution of our specific setup (280 nm full-width-at-half-maximum (FWHM)) was determined from a line scan across an edge of the MoS$_2$ flake (Figure 1e). Spatial features below the diffraction limit of the excitation wavelength can thus still be resolved, due to the nonlinear nature of the SHG process. Details of the experimental setup can be found in the Methods section and in Supplementary Figure 1.

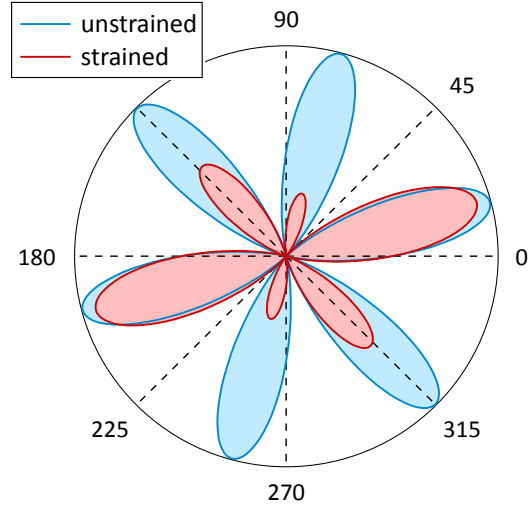

**Figure 2 | Polarization resolved SHG intensity $I_\parallel^{(2)}(2\omega)$ pattern.** The plot shows the component of the SHG signal with same polarization as the incident field. Blue line: unstrained TMD crystal; red line: with 1.0 % tensile strain in horizontal (0°) direction ($\theta = -15°$).

We will now discuss how SHG can be used to measure strain in 2D TMDs, or thin crystals in general. Polarization resolved SHG reflects the lattice symmetry of the probed crystal, where the relation between SHG intensity and crystal lattice is given by the second-order nonlinear susceptibility tensor $\chi_{ijk}^{(2)}$. Strain deforms the crystal lattice and therefore also influences $\chi_{ijk}^{(2)}$, breaking the symmetry in the SHG polarization pattern, as illustrated in Figure 2. This effect is described by an Ansatz which considers a linear strain dependence of the nonlinear susceptibility tensor [36, 37]

$$\chi_{ijk}^{(2)} = \chi_{ijk}^{(2,0)} + p_{ijklm} u_{lm}, \quad (1)$$

with $\mathbf{p}_{ijklm} = \partial \boldsymbol{\chi}_{ijk}^{(2,0)}/\partial \mathbf{u}_{lm}$. Here, $\boldsymbol{\chi}_{ijk}^{(2,0)}$ describes the second-order nonlinear susceptibility of the unstrained crystal and $\mathbf{p}_{ijklm}$ is the (fifth-rank) photoelastic tensor, which translates the strain tensor $\mathbf{u}_{lm}$ into a nonlinear susceptibility contribution. The strain tensor is symmetric ($\mathbf{u}_{lm} = \mathbf{u}_{ml}$) and the SHG process is dispersion free ($\boldsymbol{\chi}_{ijk}^{(2)} = \boldsymbol{\chi}_{ikj}^{(2)}$). Therefore, the photoelastic tensor must feature the same symmetries: $\mathbf{p}_{ijklm} = \mathbf{p}_{ikjlm} = \mathbf{p}_{ijkml} = \mathbf{p}_{ikjml}$. These symmetries reduce the number of free parameters. Moreover, depending on the crystal symmetry class, there are a set of operations under which $\mathbf{p}_{ijklm}$ is invariant. TMD monolayers have a trigonal prismatic $D_{3h}$ lattice symmetry. Considering all symmetries of the $D_{3h}$ class [40], the 2D photoelastic tensor has 12 nonzero elements, with only two free parameters $p_1$ and $p_2$:

$$p_{xxxxx} = p_1,\ p_{xxxyy} = p_2,\ p_{xyyyy} = -p_1,\ p_{xyyxx} = -p_2,$$

$$p_{yyxyy} = p_{yyxxx} = p_{yxyyy} = p_{yxyxx} = -\frac{1}{2}(p_1 + p_2),$$

$$p_{yyyyx} = p_{yyyxy} = p_{yxxyx} = p_{yxxxy} = -\frac{1}{2}(p_1 - p_2). \qquad (2)$$

Depending on the coordinate system, the strain tensor $\mathbf{u}_{lm}$ can have multiple representations. We chose a principal strain system by rotating the basis vectors by an angle $\theta$ so that the shear components vanish ($u_{xy} = u_{yx} = 0$) and chose, without loss of generality, $|u_{xx}| > |u_{yy}|$. Under the consideration of Poisson's ratio $\nu$ (the ratio of transverse to axial strain), we then obtain a 2D principal strain tensor

$$\mathbf{u}_{lm} = \begin{pmatrix} \varepsilon_{xx} - \nu\varepsilon_{yy} & 0 \\ 0 & \varepsilon_{yy} - \nu\varepsilon_{xx} \end{pmatrix}. \qquad (3)$$

After rotating the principal strain tensor back into the coordinate system of the photoelastic tensor, where the $x$-direction is defined as the armchair (AC) direction of the hexagonal crystal lattice, we can calculate the polarization resolved SHG response of strained $D_{3h}$ crystals. The induced second-order polarization is given by the nonlinear susceptibility tensor and the incident electric fields, $\mathbf{P}_i^{(2)}(2\omega) \propto \boldsymbol{\chi}_{ijk}^{(2)} \mathbf{E}_j(\omega) \mathbf{E}_k(\omega)$. In our polarization resolved SHG measurements we chose a linear polarized incident electric field under an angle $\phi$ and analyze the SHG signal with same polarization. $P_\parallel^{(2)}(2\omega)$ is the parallel polarization which corresponds to that signal and the square of it is proportional to the measured SHG intensity:

$$I_\parallel^{(2)}(2\omega) \propto \tfrac{1}{4}\bigl(A\cos(3\phi) + B\cos(2\theta+\phi)\bigr)^2, \qquad (4)$$

with $A = (1-\nu)(p_1 + p_2)(\varepsilon_{xx} + \varepsilon_{yy}) + 2\chi_0$ and $B = (1+\nu)(p_1 - p_2)(\varepsilon_{xx} - \varepsilon_{yy})$. $p_1$ and $p_2$ are the photoelastic parameters, $\varepsilon_{xx}$ and $\varepsilon_{yy}$ denote the principal strains, $\theta$ is the principal strain orientation, $\phi$ the polarization angle, and $\chi_0$ the nonlinear susceptibility parameter of the unstrained crystal lattice.

**Photoelastic tensor measurement.** The effect of strain on SHG is determined by the photoelastic parameters, which depend on the specific material. We determine these parameters for monolayer $MoS_2$ by applying different levels of uniaxial strain using a two-point bending method (Figures 3a), and measuring the polarization resolved SHG signal. For that, $MoS_2$ on a flexible substrate with length $L$ and thickness $d$ (see Methods for sample preparation) is clamped between two points and the distance $a$ between them is controlled by a motorized linear actuator. At distances $a < L$ the sample is bent, which causes tensile strain in the $MoS_2$ layer. Assuming a circular bending profile of the flexible substrate, the resulting uniaxial strain $\varepsilon_{xx}$ for a chosen distance $a$ is then given by the relation $\sin(\varepsilon_{xx} L/d) = \varepsilon_{xx} a/d$, which we solve numerically (see Supplementary Figure 3).

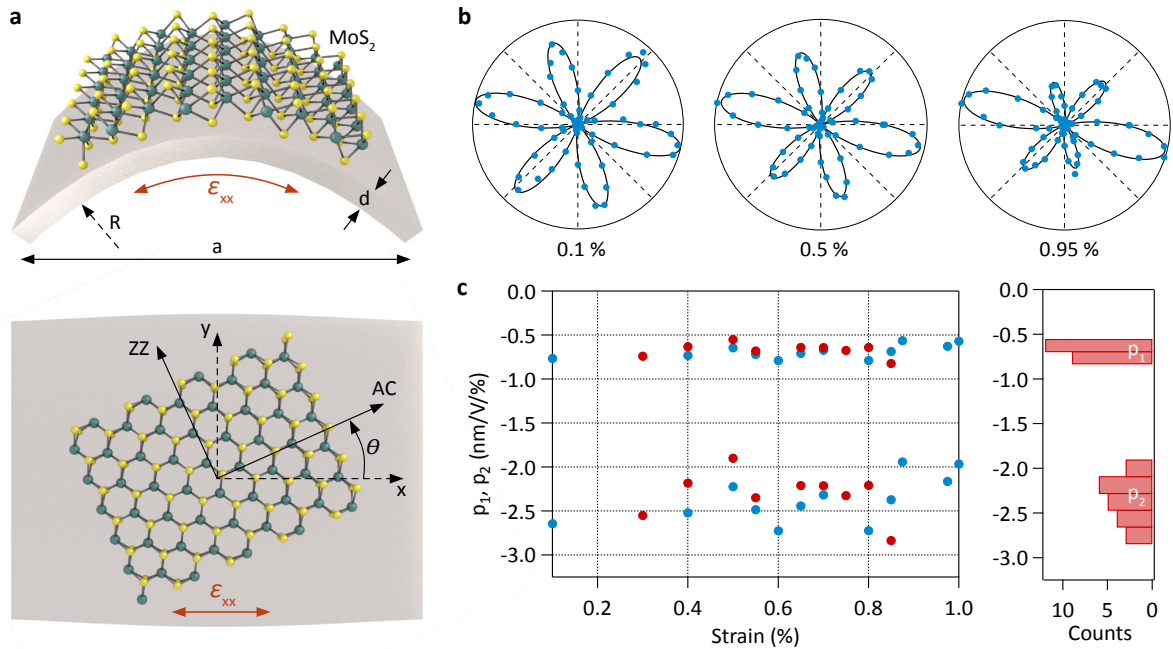

**Figure 3 | Photoelastic parameters. a,** Schematic illustration of two-point bending method (top: side view; bottom: top view). AC, armchair direction; ZZ, zig-zag direction. **b,** SHG patterns for applied tensile strains of 0.1, 0.5 and 0.95 %. (symbols: measurement data; line: fit) **c,** Photoelastic parameters $p_1$ and $p_2$ for monolayer $MoS_2$ as determined from measurements on two different

samples (plotted in different colors). Left panel: strain dependence; right panel: histogram of measurement data.

Fitting the SHG patterns at different known strain levels with equation (4), allows us to determine the photoelastic parameters $p_1$ and $p_2$. Experimentally, we find that the SHG intensity under varying uniaxial strain is constant in the uniaxial strain direction ($I_\parallel^{(2)}(\phi = \theta)$ = constant), which indicates that the two photoelastic parameters are connected by Poisson's ratio: $p_1 = \nu p_2$. This relation simplifies the nonlinear fitting procedure and, as shown in Figure 3b, the measured polarization resolved SHG data from strained MoS₂ flakes can be fitted accurately with the linear strain consideration in the nonlinear susceptibility tensor (1). In Figure 3c the photoelastic tensor parameters are plotted for two different samples for tensile strain values up to ~1 %. Both parameters are constant over this range, confirming the theoretically predicted linear relation between strain and nonlinear susceptibility (Equation (1)). Using a Poisson ratio of $\nu_{MoS2} = 0.29$ [41] and an unstrained nonlinear susceptibility of $\chi_0 = 4.5$ nm/V [32], we determine the photoelastic tensor parameters of MoS₂ monolayers as $p_1 = -0.68 \pm 0.07$ nm/V/% and $p_2 = -2.35 \pm 0.25$ nm/V/%.

**Strain imaging.** Having extracted these parameters, we can now employ SHG spectroscopy to locally probe inhomogeneous strain fields in MoS₂ monolayer samples. As discussed before, due to the nonlinear nature of SHG, the spatial resolution is higher than that obtained by linear optical strain mapping techniques, such as photoluminescence or Raman spectroscopy. However, recording a high-resolution strain map, where each strain value is determined by a full polarization scan, is extremely time consuming. We thus determine the strain from the SHG intensities at the three AC directions ($\phi = 0°, 60°, 120°$) of the MoS₂ crystal only. Using this method, we first record one complete polarization resolved measurement to determine the crystal orientation. Thereafter, for each location we acquire three SHG signals at the angles noted above, from which we determine the local strain. By taking the square root of the measured SHG intensity, being proportional to the second order polarization $P_\parallel^{(2)}(\phi) \propto \pm(I_\parallel^{(2)}(\phi))^{1/2}$, we obtain a system of equations with three unknowns – $A$, $B$, and $\theta$, presented in the Methods section. This equation system is analytically solvable and yields:

$$A = \tfrac{2}{3}(P_0 + P_{60} + P_{120}),$$

$$B = \tfrac{4}{3}(P_0^2 + P_{60}^2 + P_{120}^2 - P_0 P_{60} - P_0 P_{120} - P_{60} P_{120})^{1/2},$$

$$\theta = \tfrac{1}{2}\arctan\left(\sqrt{3}\,\frac{P_{60} - P_{120}}{2P_0 - P_{60} - P_{120}}\right), \qquad (5)$$

with $P_0 > P_{60}, P_{120}$.

$A$ and $B$ are directly related to the principal strain values $\varepsilon_{xx}$ and $\varepsilon_{yy}$ (see equation (4)), and $\theta$ is the rotation of the principal strain coordinate system relative to the crystal lattice. Uniaxial strain can then be determined as $\varepsilon_{ua} = \varepsilon_{xx} - \varepsilon_{yy}$ and biaxial strain as $\varepsilon_{bi} = \varepsilon_{yy}$, since we chose $|\varepsilon_{xx}| > |\varepsilon_{yy}|$. The relative error of the photoelastic tensor elements $p_1$ and $p_2$ is directly connected to the error of the calculated strain $\delta\varepsilon = \frac{\delta p_1}{p_1}\varepsilon = \frac{\delta p_2}{p_2}\varepsilon$. In our measurement, the relative error of the photoelastic tensor components is ~10 %, which is also the uncertainty of the measured strain values. The uncertainty of $p_1$ and $p_2$ mainly stems from our two-point bending technique, which does not fix the strained samples as good as three- or four-point bending methods. It could thus be improved by employing more sophisticated straining techniques.

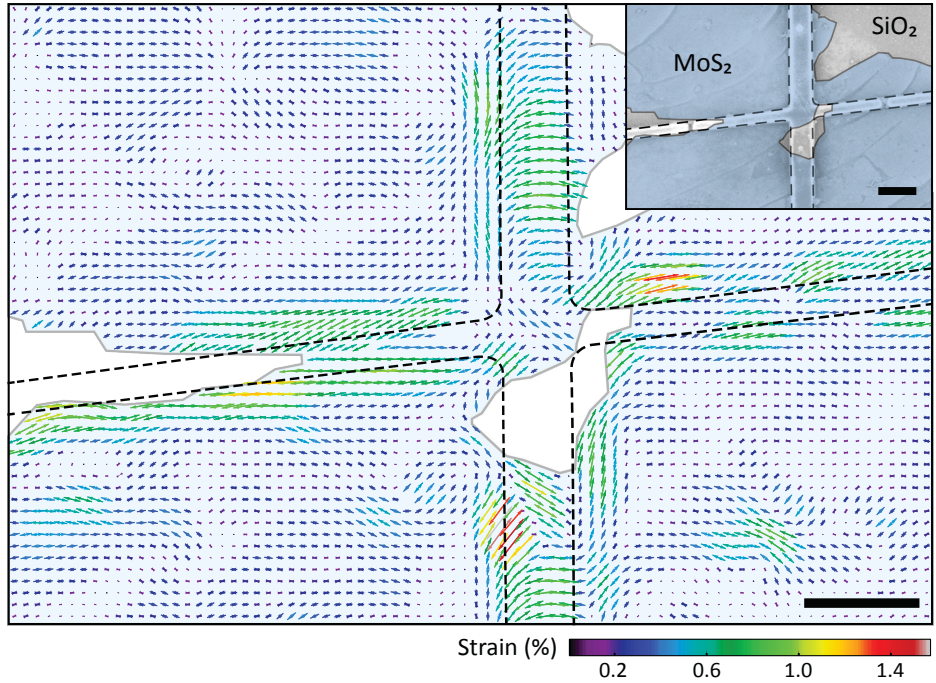

**Figure 4 | Strain imaging.** Uniaxial strain map of MoS$_2$ monolayer flake (filled color) on a lithographically defined structure (dashed lines). Arrows: measured uniaxial strain field. Inset: SEM image of sample. Scale bars, 1 µm.

We finally present an illustrative application example of the SHG strain mapping technique. Using a PC (polycarbonate) based pick-and-place technique, we transferred a mechanically exfoliated MoS$_2$ monolayer flake onto a 115 nm high lithographically defined structure on a Si/SiO$_2$ substrate, shown as scanning electron microscope (SEM) image in Figure 4 (inset). The force applied during the transfer technique causes the MoS$_2$ to be strained by the non-flat surface. A strain map is then recorded using the technique described above, and the resulting local uniaxial strains $\varepsilon_{\text{ua}}$ are plotted as vectors in Figure 4. Since all uniaxial strain values are positive we conclude that only tensile strain is present, which makes sense because the relaxed MoS$_2$ is spanned over the lithographic structure. Also contaminations on the substrate (e.g. in the lower right corner) lead to strained MoS$_2$ areas.

## DISCUSSION

In summary, we have presented a method that allows to extract the full strain tensor by means of polarization resolved SHG measurements. Using a two-point bending technique, we determined the photoelastic tensor elements for monolayer MoS$_2$. Once identified, these parameters allowed to spatially map the strain field in an inhomogeneously strained sample with 280 nm spatial resolution. Determination of the local strain tensor via three polarized SHG measurement points enables for efficient and fast strain field imaging over large sample areas, providing an unprecedented depth of information. Our method supplements and extends established optical strain measuring methods, such as Raman and photoluminescence spectroscopy. Due to the insensitivity of the SHG response to free carriers in TMD monolayers [42], our technique is less prone to artefacts that arise from local doping. Moreover, using this technique, it may be possible to image transient crystal deformations on a sub-picosecond timescale.

## METHODS

**Experimental SHG setup.** For SHG measurements we use a femtosecond Ti:sapphire laser source (200 fs, 76 MHz) that is tuned to a wavelength of 800 nm, below the MoS$_2$ band gap. A 100× confocal objective lens ($NA = 0.9$) is used to excite the sample with a diffraction-limited spot with ~1 mW average power. The SHG signal is collected in reflection geometry. In order to excite the sample with tunable linear polarization, a quarter-wave plate is first used to obtain light with circular polarization. Using a linear polarizer, mounted in a motorized rotation stage, the circularly polarized beam is then converted into a linear polarized one with polarization angle $\phi$. The same linear polarizer is used to filter out SHG light with polarization angles other than $\phi$. To suppress the excitation light, a dichroic mirror, that reflects wavelengths below 650 nm, as well as short- and band-pass filters are used, and the filtered SHG signal is detected with a sensitive photodetector. An additional beam splitter in the beam path allows to illuminate the sample with a white light source and capture the image with a CMOS camera. This allows to coarsely align the excitation laser spot on the sample. The beam splitter is removed during SHG measurements to avoid signal losses. To minimize the impact of background illumination (room light), we modulate the laser beam with a mechanical chopper and detect the photodiode signal with a lock-in amplifier. A schematic drawing of the setup is presented in Supplementary Figure 1.

**Sample preparation for two-point bending experiment.** We use polyethylene naphthalate (PEN) with a thickness of $d = 0.25$ mm and a length of $L = 18$ mm as flexible substrate. First, a layer of SU-8 (MicroChem) photoresist is spun on top of the PEN substrate and the coating is cross-linked by UV light and heat. $MoS_2$ monolayers are then selectively transferred using a pick-and-place technique. For that, $MoS_2$ is mechanically exfoliated onto a $Si/SiO_2$ wafer and a monolayer is picked up by using a Polydimethylsiloxane (PDMS) stamp covered by a polycarbonate (PC) layer. The stamp with the monolayer is then brought into contact with the SU-8 layer on the flexible substrate. After heating the sample to ~180 °C the stamp is lifted up, leaving the $MoS_2$ monolayer and the PC layer on top of the SU-8 surface. The PC layer is then dissolved with chloroform. In order to clamp down the $MoS_2$ monolayer, another SU-8 coating is spun on top. The SU-8 encapsulation is finally fully cross-linked by heating to a temperature of 200 °C for 30 minutes. This modifies the SU-8 layer such that it has the mechanical strength to reliably transfer strain from the substrate to the $MoS_2$ monolayer.

**Nonlinear polarizations along the armchair directions.**

$$P_0 = \pm P_\parallel^{(2)}(\phi = 0°) = \pm\frac{1}{2}\big(A + B\cos(2\theta)\big),$$

$$P_{60} = \mp P_\parallel^{(2)}(\phi = 60°) = \pm\frac{1}{2}\big(A - B\cos(2\theta + 60°)\big),$$

$$P_{120} = \pm P_\parallel^{(2)}(\phi = 120°) = \pm\frac{1}{2}\big(A + B\cos(2\theta + 120°)\big). \quad (6)$$

**Data availability.** The data that support the findings of this study are available from the corresponding author upon request.

**Acknowledgments:** We are grateful to Georg Reider, Juraj Darmo and Tobias Korn for helpful discussions and to Karl Unterrainer for providing access to a Ti:Sapphire laser source. We acknowledge financial support by the Austrian Science Fund FWF (START Y 539-N16) and the European Union (grant agreement No. 696656 Graphene Flagship).

**Author contributions:** L.M., M.M.F and T.M. conceived and designed the experiment. L.M., M.M.F and S.W. built the setup and carried out the measurements. M.M.F., L.M., S.W., M.P. and D.K.P. fabricated the samples. L.M. and M.M.F. analyzed the data. L.M., M.M.F. and T.M. prepared the manuscript. All authors discussed the results and commented on the manuscript.

**Competing financial interests:** The authors declare no competing financial interests.